\documentclass[aip,showpacs,onecolumn,superscriptaddress]{revtex4-1}
\usepackage{amsmath}
\usepackage{amssymb}
\usepackage{mathtext}
\usepackage{graphicx}
\usepackage{psfrag}
\usepackage{color}
\def\th{\theta}
\def\a{\alpha}

\def\lam{\lambda}
\def\eps{\epsilon}
\def\part{\partial}

\begin{document}

\title{
Solitary phase waves in a chain of autonomous oscillators}
\author{Philip Rosenau}
\affiliation{Department of Mathematics, Tel-Aviv University\\
 Tel-Aviv 69978, Israel}
\author{Arkady Pikovsky}
\affiliation{Institute for Physics and Astronomy,
University of Potsdam, Karl-Liebknecht-Strasse 24/25, 14476 Potsdam-Golm, Germany}
\affiliation{National Research University, Higher School of Economics,
Bolshaya Pecherskaya St. 25/12, 603155 Nizhny Novgorod, Russia}
\date{\today}

\begin{abstract}
In the present paper we study phase waves of self-sustained oscillators with a nearest
neighbor dispersive coupling on an infinite lattice. To analyze the underlying dynamics we approximate the lattice with a  quasi-continuum, QC. The resulting partial differential model is then further reduced to the Gardner equation which predicts many properties of the underlying solitary structures. Using an iterative procedure on the original lattice equations we determine the shapes of solitary waves, kinks, and the flat-like solitons, that we refer to as flatons. Direct numerical experiments reveal that the interaction of solitons and flatons on the lattice is notably clean. All in all we find that both the QC and the Gardner equation predict remarkably well the discrete  patterns and their dynamics.
\end{abstract}

%\pacs{05.45.Xt,    % Synchronization; coupled oscillators
%      05.40.-a,    % Fluctuation phenomena, random processes, noise, and Brownian motion
 %     02.50.Ey     % Stochastic processes
%}
\maketitle

\begin{quotation}
Interacting limit cycle oscillators play a fundamental role in synchronization studies. When the coupling is small,
 systems dynamics reduces to that of the oscillators phases.
In many set-ups, as in the seminal Kuramoto model, the
interaction is dissipative and leads ultimately to  synchrony of all phases.
Yet, in many experimental set-ups the coupling
is dispersive  with the resulting phase equations being conservative.
We focus on a conservative phase dynamics
 on a one-dimensional lattice and demonstrate the existence of a very robust dynamics of solitary waves. A crucial role in the understanding
of the dynamics is played by its quasi-continuum approximation via a partial differential equation which provides a remarkably
accurate description of the underlying phenomena on the lattice.
\end{quotation}

\section{Introduction}
Dynamics of networks of oscillators have gathered  in recent years  considerable attention. The dynamics of even the simplest
network architectures, like global coupling in a population~\cite{Pikovsky-Rosenblum-15} or
a local coupling on a regular lattice (see e.g.,~\cite{Yanchuk-Wolfrum-08}),
 is highly nontrivial even for the simplest Kuramoto-Sakaguchi type of
interactions~\cite{Sakaguchi-Kuramoto-86}. In the latter case, one addresses the phase dynamics
of oscillators coupled via their first harmonics, with an additional phase shift. This
phase shift determines the relative importance of dissipative, diffusion-type, and conservative
(dispersive) interactions. For the diffusion-type coupling, the interaction results in a global
synchronization of a homogeneous lattice (in a inhomogeneous lattice with random oscillator
frequencies, the diffusive coupling should be strong enough to ensure
synchrony~\cite{Ermentrout-Kopell-84}.)  Homogeneous oscillator
lattices with a  purely conservative coupling follow a very different path for their phase dynamics, which is, surprisingly enough, Hamiltonian
leading to a formation of a nontrivial waves, like compactons or kovatons ~\cite{Rosenau-Pikovsky-05,Pikovsky-Rosenau-06,Ahnert-Pikovsky-08,Rosenau_2018}.

In the present paper we extend our previous work and address  both analytically and numerically
a dispersive variant of the
Kuramoto-Sagakuchi chain. Studying the solitary structures we find solitons in a bounded range of velocities. At range's edge solitons collapse and kink/anti-kink emerge. However, close to the transition's threshold, we find a narrow strip of velocities wherein solitons undergo a structural change and rather than grow with amplitude they widen and turn into flat-top solitons, referred to as flatons. Notably, interaction between solitons and flatons is found to be remarkably clean. A purely
dispersive coupling   of self-sustained oscillators, of the type studied here, is relevant in micromechanical
oscillators studied both theoretically~\cite{Cross_etal-04,Cross_etal-06,Lee-Cross-11,Matheny_etal-14,Fon_eal-17} and
more recently were explored experimentally
 \cite{Matheny_etal-19}. Note that the long chains we focus on are not easily
 accessible experimentally. The more realistic short chains where boundary effects matter
  will be a subject of future studies.

\section{The basic model}
Consider a chain of self-sustained, autonomous, oscillators with a nearest-neighbor
coupling, described via their complex amplitudes
$A_n$:
\begin{equation}
\frac{d A_n}{dt}=i\omega A_n+ \mu A_n(1-|A_n|^2)+i\varepsilon (A_{n-1}-2A_n+A_{n+1})\;.
\label{eq:cgll}
\end{equation}
The amplitudes in \eqref{eq:cgll} were normalized with the equilibrium amplitude
of a single oscillator being unity, whereas $\mu$, assumed to be large, governs the relaxation rate to the equilibrium,
 so that the limit cycle oscillations are strongly stable. Contrary to the large dissipation
of the local amplitude dynamics, the present coupling is assumed to be purely conservative as found in
  nano-electro-mechanical setups addressed in \cite{Cross_etal-04,Cross_etal-06,Lee-Cross-11,Matheny_etal-14,Fon_eal-17} and
recently realized experimentally
 \cite{Matheny_etal-19}. In the  $\mu\to\infty$ limit one may neglect the changes in
 amplitude's modulus and set $A_n=e^{i\varphi_n}$, where $\varphi_n$ is the phase of the oscillator. This leads
to a phase chain model
\begin{equation}
\frac{d\varphi_n}{dt}=\omega +\varepsilon \Big(\cos(\varphi_{n+1}-\varphi_n)+
\cos(\varphi_{n-1}-\varphi_n)-2\Big)\;.
\label{eq:plm1}
\end{equation}
In chain \eqref{eq:plm1}, any linear phase profile $\varphi_n=(\pi/2-\alpha) n$ is uniformly rotating (a so-called twisted state~\cite{Wiley-etal-06}).
 To study the deviations from this plane wave we introduce the phase difference  $\theta_n=\varphi_{n+1}-\varphi_n+\alpha-\pi/2$,
and rescale the time $\varepsilon t\to t$, to obtain
the following basic model
\begin{equation}
\frac{d\th_n}{dt}=\sin(\alpha-\theta_{n+1})-\sin(\alpha-\theta_{n-1}).
\label{eq:plm3}
\end{equation}
 As  is clear from its derivation, Eq.~\eqref{eq:plm3} which is the basis of our studies, describes phase waves on the top of
the plane wave, whereas $\alpha$ defines the slope of the background linear
phase profile. The very particular  $\alpha=\pi/2$ case was addressed in Refs.~[\onlinecite{Rosenau-Pikovsky-05,Pikovsky-Rosenau-06}].

Noteworthy are the invariance properties of Eq.~\eqref{eq:plm3} under $(\theta,\alpha) \to -(\theta,\alpha)$ and
$(\theta,\alpha,n)\to (-\theta,\pi-\alpha,-n)$ and especially its invariance   under
\begin{equation}
\theta\to 2\alpha-\theta.
\label{eq:Invariance}
\end{equation}

Consequently both $\th=0$ and $\th=2\a$ are solutions as is $\th=\a$ which is mapped into itself. \\

Among the features of Eq.~\eqref{eq:plm3} we note the
 dispersion relation $\omega=2\cos\alpha\sin k$, $-\pi\leq k\leq \pi$ of its linear waves and the aforementioned "sonic vacuum" when $\alpha=\pi/2$~\cite{Rosenau-Pikovsky-05,Pikovsky-Rosenau-06}, for
linear waves are absent and the lattice becomes essentially nonlinear. Finally, we note the conservation laws
\begin{equation}
I_{1}=\sum_{n}\th_{n},~~~I_{2}=\sum_{n}\cos(\a-\th_{n}),~~~\text{and}~~ I_{3}=\sum_{n}(-1)^{n}\th_{n},
\label{eq:Conservation}
\end{equation}
valid on an infinite chain.

\section{Waves in a Quasi-continuum}
\subsection{General Features}

In spite of their innocuous appearance, Eqs.~\eqref{eq:plm3} describe a
complicated nonlinear system which defies a direct analysis.
As in previous works \cite{Rosenau-Pikovsky-05,Pikovsky-Rosenau-06,Rosenau_2018},
to gain insight into its dynamics we shall adopt a quasi-continuous description
wherein the discrete system is replaced with a continuous formulation
which keeps a trace of its discrete origin.
To this  end we approximate
the chain as:
$\theta_{j}(t)\to \theta(x,t)$,  and $f_{j+1}-f_{j-1}\to 2(f_{x}+ \frac{1}{6} f_{xxx})$
which yields
\begin{equation}
\frac{1}{2}\frac{\partial{\th}}{\partial{t}} + \Big(\frac{\partial}{\partial{x}}+\frac{1}{6}\frac{\partial^{3}}{\partial{{x^3}}}\Big)\sin({\th}-\a) =0.
\label{eq:qc1}
\end{equation}
As there is no small parameter in the problem, Eq.~\eqref{eq:qc1} cannot, in the strict mathematical sense,
be considered  as an asymptotic description of the discrete problem and its utility can be judged only
posteriori. Nonetheless both
in the present problem and in a large variety of other problems, cf.~\cite{Rosenau-Pikovsky-05,Pikovsky-Rosenau-06,Rosenau_2018},
it captures both the qualitative and quantitative properties of the discrete solitary waves remarkably well.

Similarly to the discrete system, Eq.~\eqref{eq:qc1} is also invariant under
$\th \to 2\alpha - \th$. Thus, if $\th_{1}$
is solution so is $\th_{2}=2\alpha-\th_{1}$.
This is a cyclic property with $\th_{2}$ leading back to $\th_{1}$ and as in the discrete case the trivial $\th=0$ solution yields $\th=2\alpha$ whereas $\th=\a$ is mapped into itself.

Separating the linear convection, we rewrite \eqref{eq:qc1} as
 \begin{equation}
\frac{1}{2} \th_{t} + \cos{\a}\th_{x} + C_{qc}(\th)\th_{x} + \frac{1}{6} \frac{\partial^{3}}{\partial x^{3}}
\sin(\th-\a)=0\;,
\label{eq:qc2}
\end{equation}
where
\begin{equation}
  C_{qc}(\th)= 2\sin(\frac{\th}{2})\sin(\a-\frac{\th}{2})\;.
\end{equation}
 Note the non-monotone nature of $C_{qc}(\th)$;  it attains its maximal
 value at $\th=\a$ and vanishes at both $\th=0$ and $\th=2\a$, which
 merely reflects its  invariance under \eqref{eq:Invariance}.

 Eq.~\eqref{eq:qc1} conserves four local quantities
\begin{equation}
I_{1}=\int{\th}dx,~~~ I_{2}=\int{Q(\th)}dx, ~~~,I_{3}=\int{\sin{\sqrt{6}(x+x_{0})\th}}dx,~~~
I_{4}=\int{\cos{\sqrt{6}(x+x_{0})\th}}dx\;,
\end{equation}
where $Q(\th)=\int{\sin(\th'-\a)}d\theta'=\cos(\alpha-\th)$. $I_{1}$ and $I_{2}$  are in a direct correspondence
with the corresponding discrete conservation quantities. Notably, the QC admits also a Lagrangian (for more details see ~\cite {Rosenau_2018})

\begin{equation}
{\cal{L}} =\int \int \Big[\frac{1}{2}\psi_{x}\psi_{t}- Q\Big({\bf{L}}\psi_{x}\Big)\Big]dxdt;,
\end{equation}
where
\begin{equation}
{\bf{L}}=\sqrt{1+\partial^{2}_{x}}, ~~~ \th= {\bf{L}}{v}, ~~~\text{and}~~~v=\psi_{x}.
\end{equation}
Consequently, the QC  conserves also the momentum $\int{v^{2}}dx$, and in the original variables

\begin{equation}
I_{5} =\int{\th {\bf{L}}^{-2}\th}dx\;.
\end{equation}

\subsection { The Gardner Approximation }
To unfold the key phenomena we begin with a weakly non-linear regime, $\th<<1$, wherein
\begin{equation}
\frac{1}{2}\th_{t}+\cos{\a}\th_{x} +(\frac{\sin{\a}}{2}\th^{2}-\frac{\cos{\a}}{6}\th^{3})_{x} +
\frac{\cos{\a}}{6}\frac{\partial^{3} \th}{\partial x^{3}} =0\;.
\label{eq:qc3}
\end{equation}
Note that whereas, on one hand, we have neglected the nonlinear corrections to the third derivative which, insofar that $\th -\a \neq \pm \pi/2$, has only a minor quantitative impact,
on the other hand, to preserve the crucial non-monotone nature of convection we have carried its expansion to the third order.
Using the Galilean invariance to dispense with convection's
linear part, after normalization  Eq.~\eqref{eq:qc3} begets the celebrated Gardner equation
\begin{equation}
u_t +C_{G}(u)u_x+u_{xxx}=0\;,~~~~\text{where}~~C_{G}=6u(1-u).
\label{eq:G1}
\end{equation}
 Both $C_{G}$ and  Eq.~\eqref{eq:G1} are invariant under $u \rightarrow 1-u$, which echoes the invariance of the original lattice and its QC rendition under \eqref{eq:Invariance}.  Its solitons, traveling with speed $\lambda$, satisfy an ordinary differential equation
with respect to $s=x-\lambda t$:
\begin{equation}
\frac{1}{2}u_s^2 + P_{G}(u)=0\;,~~~{\rm where}~~2P_{G}(u)=- \lambda u^2 + 2u^3 - u^4\;.
\label{eq:G2}
\end{equation}
Note that due to the defocusing effect of the cubic term,
the potential peaks at $u=\frac{3}{4}[1+\sqrt{1-8\lam/9}]$, and comes down as the speed increases.  Consequently,
the resulting solitons
\begin{equation}
u=\frac{\lambda}{1+ \sqrt{1-\lambda}
\cosh{\left(\sqrt{\lambda} (x- \lambda t)\right)}}
\label{eq:TW2modkdv}
\end{equation}
have a bounded range  of admissible
propagation speeds: $0<\lambda<1$. At the limiting velocity
$\lambda =1$, potential's peaks touches the $u$-axis and
the soliton solution~\eqref{eq:TW2modkdv} flattens into a constant $=1$.
This is a singular limit at  which both kink and an anti-kink form
\begin{equation}
u=\frac{1}{1+ \exp(\mp s)}\;,~~~
{\rm{where}}~~~ s=x-t\;.
\label{gardner2}
\end{equation}

Close to the edge of solitons upper velocity range, there is a narrow strip of velocities where solitons undergo
a structural change  and rather than grow with amplitude
they begin to widen and their top flattens. To extract these features from Eq.~\eqref{eq:TW2modkdv},
let
$$ \lambda_{f} = 1-\eps^{2}\;,~~~~
{\rm{where}}~~ 0<\epsilon \ll 1,$$
to obtain
\begin{equation}
u=  \frac{1-\eps^{2}}{1+\eps\cosh{\left[\sqrt{\lambda_f}
(x-  \lambda_{f}t)\right]}},
\label{ueps}
\end{equation}
with soliton's amplitude being $u_{max} =1-\eps$.
The extent of soliton's  widening
is expressed via $x_{1/2}$, where soliton's amplitude
has decreased by half;
\begin{equation}
x_{1/2} \simeq \ln{\frac{2}{\eps}}.
\label{x0eps}
\end{equation}

Thus, amplitude(velocity) changes $\sim 1-\epsilon$ ($\sim 1-\epsilon^{2}$) which are
 pretty much numerically unobservable, cause
 solitons to widen as $\sim \ln{1/\eps}$.
 We shall refer to the flat-like solitons as {\it flatons}.

The proximity of flatons velocities to the edge of the admissible speeds range enables to approximate them by a kink-antikink pair
 placed at $2x_{1/2} \gg 1$ from each other:
\begin{equation}
u\cong\frac{1}{1 + \exp{\left(|x|-x_{1/2}\right)}}, ~~x\in (-\infty,\infty),
\label{flaton}
\end{equation}
and provides an upper bound to all flatons.

\subsection{Analysis of the Traveling Waves}
\label{sec:twfna}

\begin{figure}[ht]
\includegraphics[width=0.5\columnwidth]{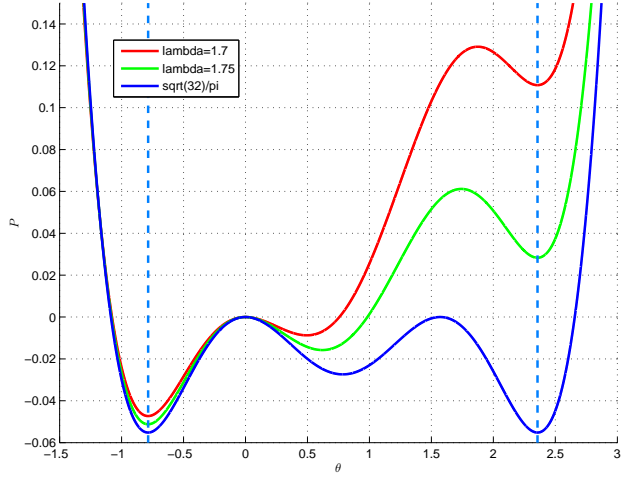}
\caption{The potential landscape for $\a=\pi/4$, with $(\lam_{min},\lam_{max})=
(\sqrt 2, 4\sqrt{2}/\pi)$. The three plots display the potential corresponding to
 $\lam = 1.7, 1.75~$ and $~4\sqrt{2}/\pi=\lambda_{max}$.
At the critical
speed  $ \lam_{max}$ potential's positive peak touches the $\th$-axis,
soliton dissolves and kink/antikink connects the two peaks. The two vertical
$\th=\alpha \pm \pi/2$ lines where  Eq.~\eqref{eq:twqc} becomes singular, bound the admissible domain.}
 \label{fig:pot}
\end{figure}

We now proceed to unfold the solitary wave structure of QC, Eq.~\eqref{eq:qc1}.
Seeking travelling waves $\theta=\theta(s=x-\lam t)$, upon one integration we have
\begin{equation}
-\frac{\lam}{2}\th + \sin(\th-\a)+\sin{\a}+ \frac{1}{6}\sin(\th-\a)'' =0,
\label{eq:TW1}
\end{equation}
and, as their small amplitude regime indicates, traveling waves call for $2\cos{\a}<\lam$.
 Integrating  Eq.~\eqref{eq:TW1} we have
\begin{equation}
\frac{1}{6}\cos^{2}(\th-\a)\th_{s}^{2}+P_{qc}(\lam,\a;\th)=0\;,
\label{eq:twqc}
\end{equation}
where the potential $P_{qc}$ reads
\begin{equation}
P_{qc}(\lam,\a;\th) = -\lam \Big[\th\sin(\th-\a)+\cos(\th-\a)-\cos\alpha]+
\Big[\sin(\th-\a)-\sin\alpha\Big]^{2}\;.
\label{P_{qc}}
\end{equation}
  A typical potential landscape  for $\a=\pi/4$ is displayed in Fig.~\ref{fig:pot}
  for three values of $\lam$.
  As in the weakly nonlinear case, the bounded potential sets an upper bound at which propagation is possible,
corresponding to  potential's top descending  toward the $\th$-axis at $\th=2\a$ with $\lam=2\sin{\a}/\a$, where the soliton flattens into a constant and kink/anti-kink emerge. Consequently,
\begin{equation}
2\cos\a=\lam_{min}<\lam <\lam_{max} = 2\frac{\sin{\a}}{\a}
%\frac{\big(1+\sin\a\big)^{2}}{0.5\pi+\a-\cos\a}.
\end{equation}
 determines the interval of admissible velocities of solitary waves.

\begin{figure}[!hbt]
\centering
\psfrag{xlabel}[cc][cc]{space}
\psfrag{ylabel}[cc][cc]{field $\theta/\pi$}
\psfrag{a}[cc][cc]{(a)}
\includegraphics[width=0.45\textwidth]{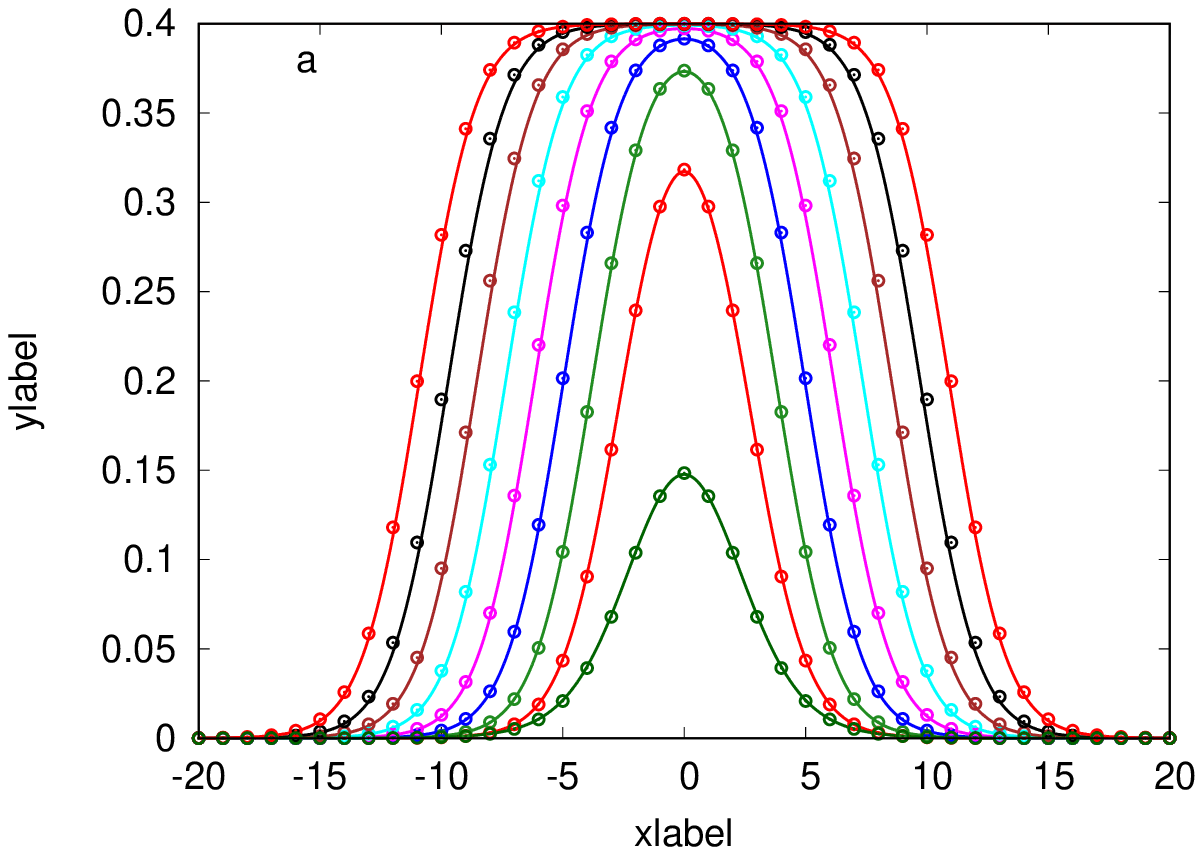}
\psfrag{xlabel1}[cc][cc]{amplitude $\theta_{max}/\pi$}
\psfrag{ylabel1}[cc][cc]{velocity}
\psfrag{b}[cc][cc]{(b)}
\includegraphics[width=0.45\textwidth]{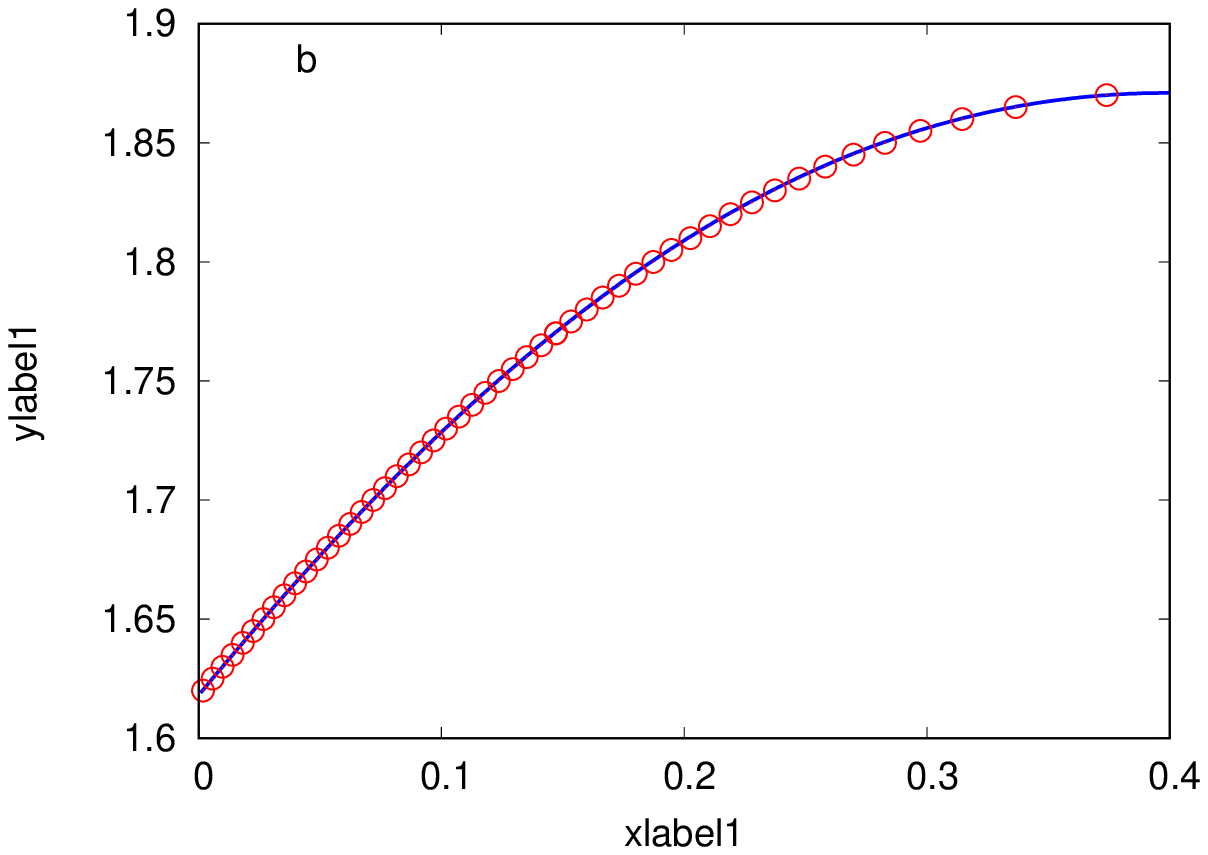}\hfill
\caption{Waves for $\alpha=0.2\pi$. (a): Solitary solutions of the chain Eq.~\eqref{it5} (circles) superimposed on
 the QC rendition, Eq.~\eqref{eq:twqc}, for different  values of their velocity deviations from its upper bound. From bottom to top: $\lambda_{max}-\lambda= 10^{-m}$, $m=1...9$.
 (b): Velocity of solitons vs. their amplitude in both discrete and QC renditions. Observe the remarkable fit between the QC and its discrete antecedent.}
\label{fig:swl}
\end{figure}
%.0.1,\;10^{-2}\;,10^{-3},\;
%10^{-4},\;10^{-5},\; 10^{-6},\; 10^{-7},\; 10^{-8},\; 10^{-9}$.

 On the basis of the weakly nonlinear regime,
 we anticipate that as $\lambda \to \lambda_{max}=2\sin(\a)/\a$,
 solitons turn into flatons which,  as illustrated in Fig.~\ref{fig:swl}, is indeed the case.

Note  the structural singularity of Eq.~\eqref{eq:twqc} at
$\th - \a =\pm\pi/2$, denoted by the vertical lines on the potential landscape, where Eqs.~\eqref{eq:qc1} and \eqref{P_{qc}} degenerate, setting $\a=\pi/2$ as the highest admissible value of parameter $\a$ with the corresponding maximal amplitude
$\th=2\a=\pi$, and the maximal speed of the kink being $\lam_{max}=\frac{4}{\pi}$.
In the special $\a=\pi/2$  case, rather than  a sequence of flatons approaching the kink limit,
a kovaton emerges which, since Eq.~\eqref{eq:qc1} becomes
singular both  at $\th=0$ and at the top $\th=\pi$, is and strictly  compact there
(see~\cite{Rosenau-Pikovsky-05,Pikovsky-Rosenau-06} for a full elaboration of this case).
Also, unlike flatons where every width corresponds to a different speed, however minutely different,
 {\it all kovatons travel at exactly the same limiting velocity of their corresponding kink} with their width being  chosen at will.

\section{Traveling waves on the chain}
 Our starting point is the original chain equation \eqref{eq:plm3} rewritten as
\begin{equation}
\dot{\theta}_n(t)=\sin\alpha\Big(\cos \theta_{n+1}(t)-\cos \theta_{n-1}(t)\Big)-
\cos\alpha\Big(\sin \theta_{n+1}-\sin \theta_{n-1}\Big)\;.
\label{it1}
\end{equation}
Seeking travelling waves $\theta_n(t)=\Theta(t-an)$ of Eq.\eqref{it1}
 where $a=1/\lambda$ is the inverse velocity, we obtain an advance-delay equation
\begin{equation}
\dot{\Theta}=\sin\alpha \Big(\cos \Theta(t-a)-\cos \Theta(t+a)\Big)-\cos\alpha\Big(\sin \Theta(t-a)-\sin \Theta (t+a)\Big).
\label{it2}
\end{equation}
We integrate Eq.~\eqref{it2} once to obtain
\begin{equation}
\Theta(t)=\int_{t-a}^{t+a} \Big[\sin\alpha (1-\cos \Theta(x))+\cos\alpha \sin \Theta(x)\Big] dx,
\label{it3}
\end{equation}
with the integration constant chosen to assure that $\Theta=0$
is a solution. In what follows \eqref{it3} will be a starting point for the following iterative procedure.

\subsection{Kinks}

Assume that there is a kink connecting $\Theta=0$ with $\Theta=\Theta_0$.
Setting $\Theta=\Theta_0$ in \eqref{it3}, begets a condition relating $a$ with $\Theta_0$:
\begin{equation}
\Theta_0=2a[\sin\alpha-\sin(\alpha-\Theta_0)]\;.
\label{eq:it4}
\end{equation}
However, symmetry \eqref{eq:Invariance} dictates that  for a given inverse velocity $a$, there should be a solution
connecting $\Theta_1=2\alpha$ with $\Theta_2=2\alpha-\Theta_0$.
This leads to an additional condition relating $a$ and $\Theta_0$:
\begin{gather}
2\alpha=2a[\sin\alpha-\sin(\alpha-2\alpha)]=4a\sin\alpha\;,\label{eq:it6}\\
2\alpha-\Theta_0=2a[\sin\alpha-\sin(\alpha-2\alpha+\Theta_0)]=
2a[\sin\alpha-\sin(\Theta_0-\alpha)]\label{eq:it7}\;.
\end{gather}
Adding \eqref{eq:it4} and \eqref{eq:it7}, we have
\[
2\alpha=4a\sin\alpha\;,
\]
which coincides with the $\lambda_{max}$  derived in Section~\ref{sec:twfna}.
Using $a=\frac{\alpha}{2\sin\alpha}$ in \eqref{eq:it7},
we have
\[
\frac{\sin(\Theta_0-\alpha)}{\Theta_0-\alpha}=\frac{\sin\alpha}{\alpha},
\]
with the obvious solution $\Theta_0=2\alpha$, which may serve as kink's amplitude.

\begin{figure}[!hbt]
\centering
\psfrag{xlabel}[cc][cc]{space}
\psfrag{ylabel}[cc][cc]{field $\Theta/\pi$}
\includegraphics[width=0.7\textwidth]{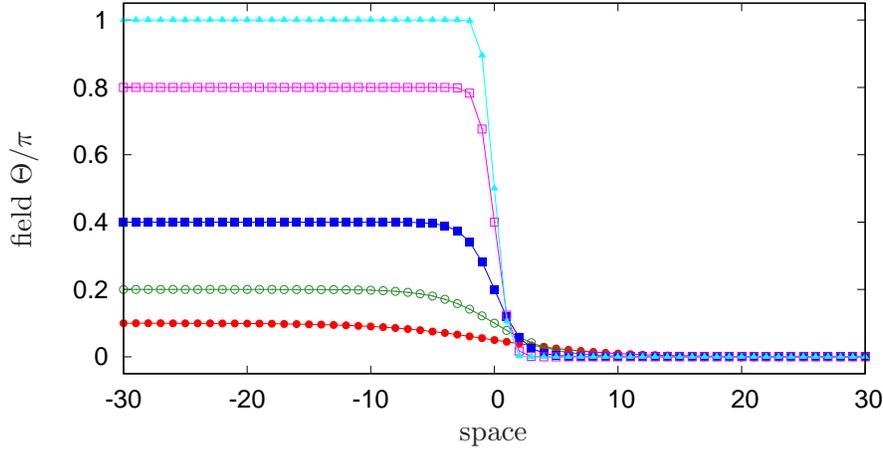}
\caption{Discrete kinks for different values of $\alpha$: from bottom to top
$\alpha=0.05\pi,\;0.1\pi,\;0.2\pi,\;0.4\pi$, and $0.5\pi$. In the last, exceptional, case the tail decays at doubly exponential rate.}
\label{fig2-k}
\end{figure}

To determine the kink we solve Eq.~\eqref{it3}
 iteratively:
\begin{equation}
\Theta^{(k+1)}(t)=\int_{t-a}^{t+a} [\sin\alpha (1-\cos \Theta^{(k)}(x))+\cos\alpha \sin \Theta^{(k)}(x)]  dx.
\label{it4}
\end{equation}
Starting from an initial ansatz $\Theta^{(0)}(x)$, having a proper asymptotic behavior at $x\to\pm\infty$,
these iterations converge and yield the kink profiles
shown in Fig.~\ref{fig2-k}.

\subsection{Solitary Waves}
 In order to apply the iterative procedure  due to V. Petviashvili~\cite{Petviashvili-76,Petviashvili-81} to solitary waves, to avoid convergence to the trivial solution we modify it
 by introducing an intermediate  normalization step

\begin{equation}
\tilde{\Theta}(t)=\int_{t-a}^{t+a}  [\sin\alpha (1-\cos \Theta^{(k)}(x))+\cos\alpha \sin \Theta^{(k)}(x)]
dx,\qquad \Theta^{(k+1)}=
\left(\frac{||\Theta^{(k)}||}{||\tilde{\Theta}||}\right)^{\gamma}\tilde{\Theta},
\label{it5}
\end{equation}
where $||\diamond||$ stands for any norm (in our implementation the $L_1$-norm was used).
Also, to assure a faster convergence the exponent $1<\gamma\leq 2$ was adjusted to the assumed $\alpha$ and
 $a$ (though the convergence itself does not depend  on $\gamma$).
When carrying the iterations \eqref{it5},
we fix $\alpha$ and $a$
and start with a solitary profile. The resulting iterations yield a solitary solution on the chain.
In Fig.~\ref{fig:swl} we compare the discrete solitary solutions with the
corresponding QC solitary solutions obtained solving Eq.~\eqref{eq:twqc}
and, as clearly seen, find a remarkable overlap attesting to the utility of the QC rendition.

\begin{figure}[!hbt]
\centering
\psfrag{xlabel}[cc][cc]{space}
\psfrag{ylabel}[cc][cc]{time}
\psfrag{xlabel1}[cc][cc]{space}
\psfrag{ylabel1}[cc][cc]{time}
\psfrag{a}[cc][cc]{(a)}
\psfrag{b}[cc][cc]{(b)}
\psfrag{p0}[cc][cc]{$0$}
\psfrag{p1}[cc][cc]{$\pi$}
\psfrag{p2}[cc][cc]{$2\pi$}
%\psfrag{f}[cc][cc]{(f)}
\includegraphics[width=0.48\textwidth]{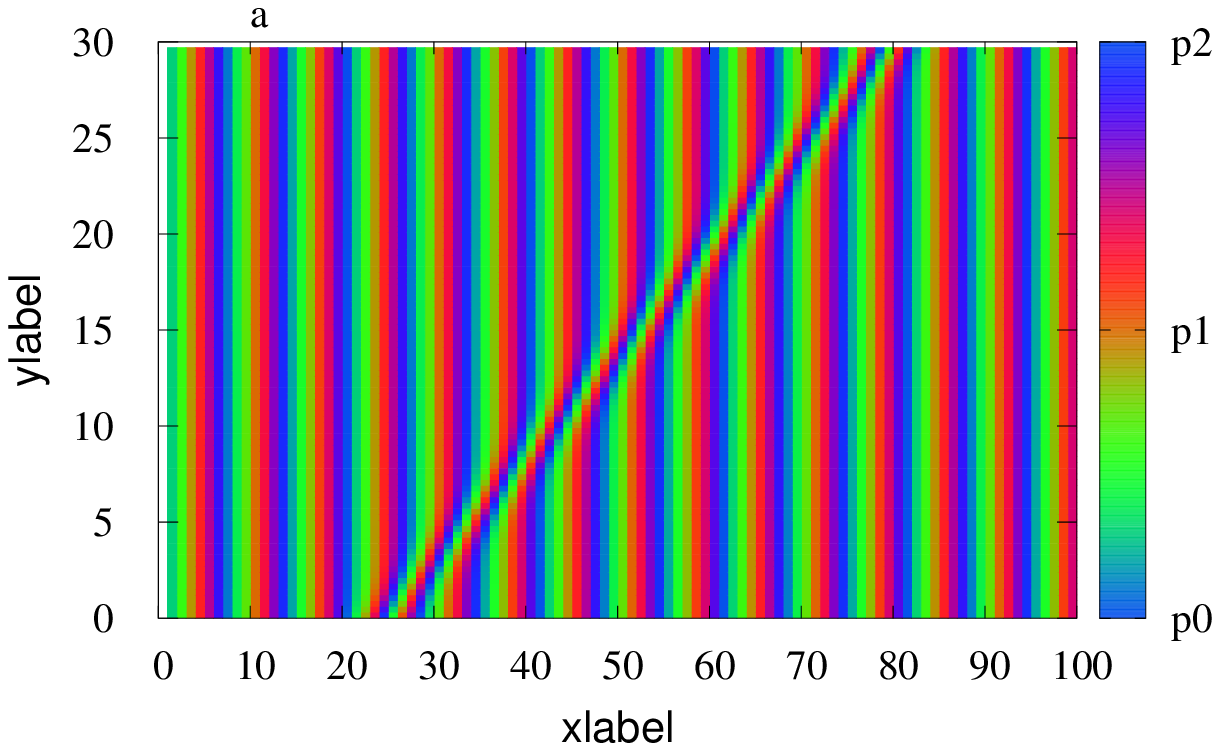}\hfill
\includegraphics[width=0.48\textwidth]{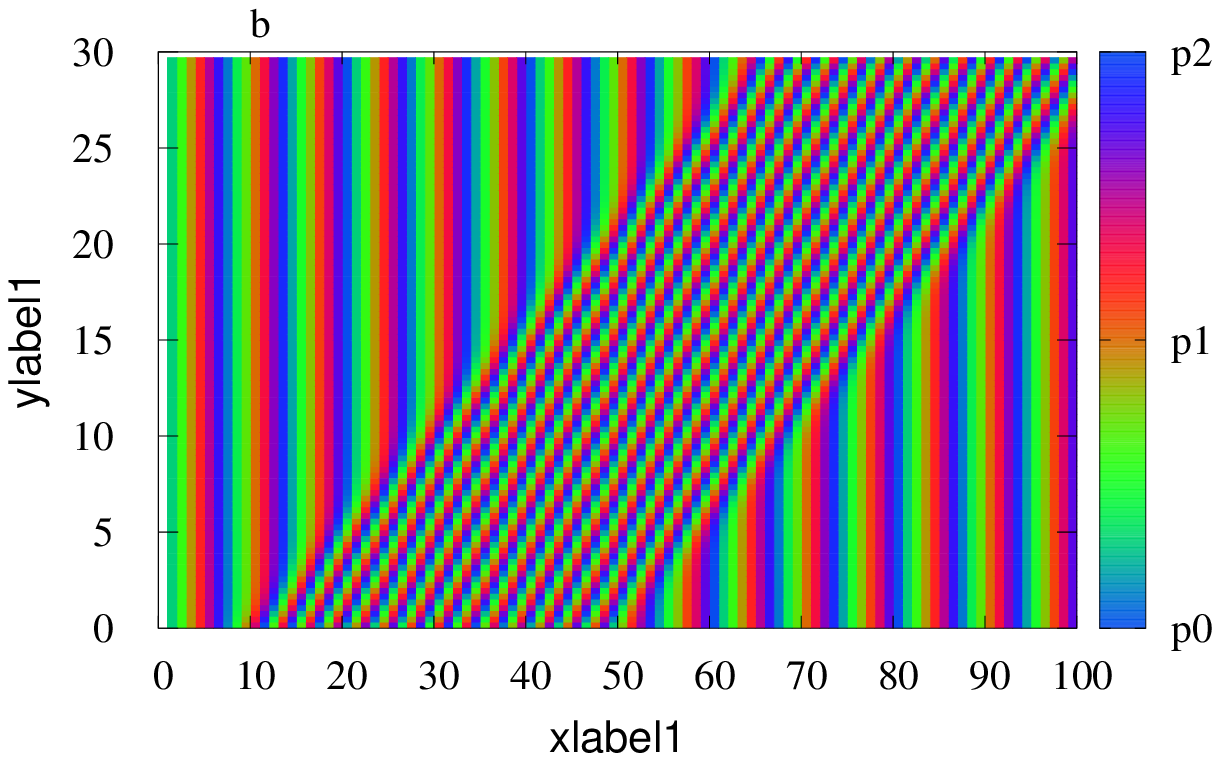}
\caption{Space-time plots of a soliton (panel (a)) and of a flaton (panel (b))
in terms of the phases $\varphi_n(t)$ (color-coded values), for $\alpha=0.2\pi$. The background stripes
represent the plane wave
on base of which solitary waves propagate.}
\label{fig:swph}
\end{figure}

It is instructive to represent the solitary waves in terms of the original
phases $\varphi_n$ rather than in the phase differences $\theta_n$ (cf. Eqs.~\eqref{eq:plm1},\eqref{eq:plm3}).
This is done in Fig.~\ref{fig:swph}, where we have adopted the reference frame with $\omega=0$.

\section{Direct simulations of the chain}
We now proceed to present the results of our  direct simulations of the chain~\eqref{it1}.
We address two basic initial-value problems.

\subsection{Evolution of an initial step}

\begin{figure}[!hbt]
\centering
\psfrag{xlabel}[cc][cc]{space}
\psfrag{ylabel}[cc][cc]{filed $\theta_n/\pi$}
\psfrag{a}[cc][cc]{(a)}
\psfrag{b}[cc][cc]{(b)}
\psfrag{c}[cc][cc]{(c)}
\psfrag{d}[cc][cc]{(d)}
\psfrag{e}[cc][cc]{(e)}
\psfrag{f}[cc][cc]{(f)}
\includegraphics[width=0.48\textwidth]{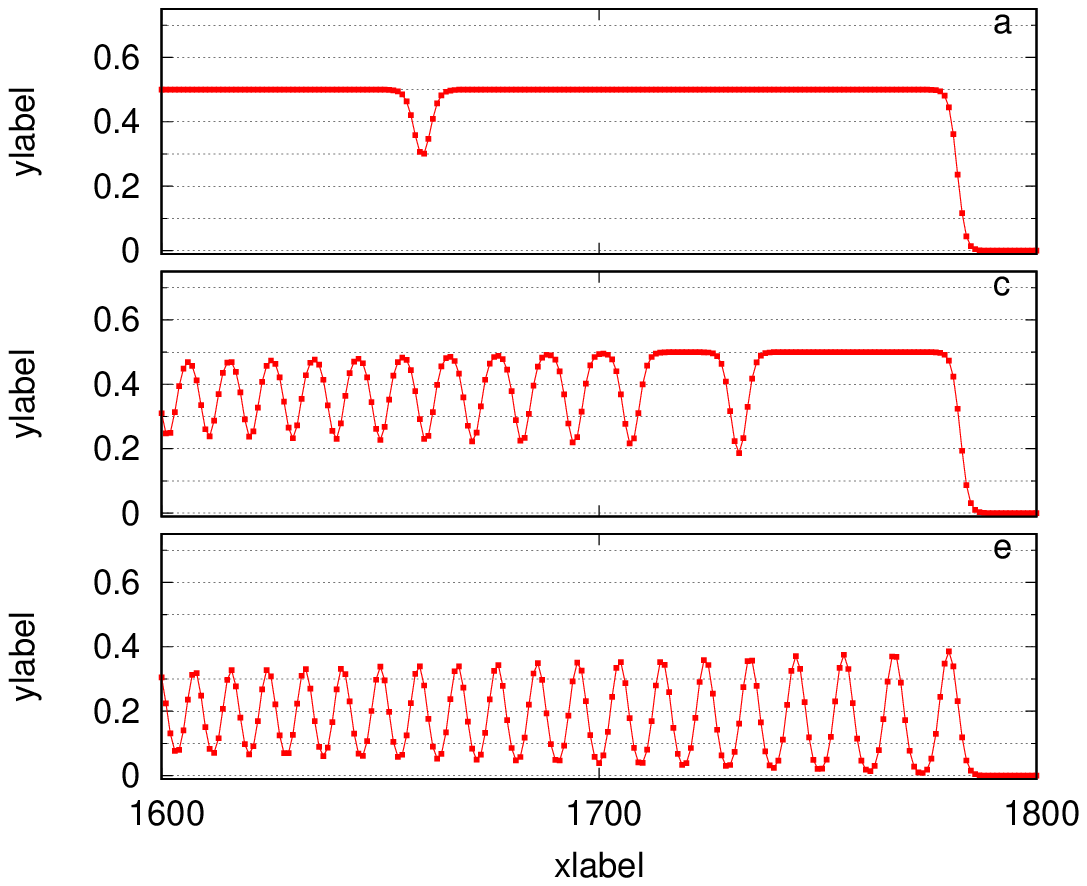}\hfill
\includegraphics[width=0.48\textwidth]{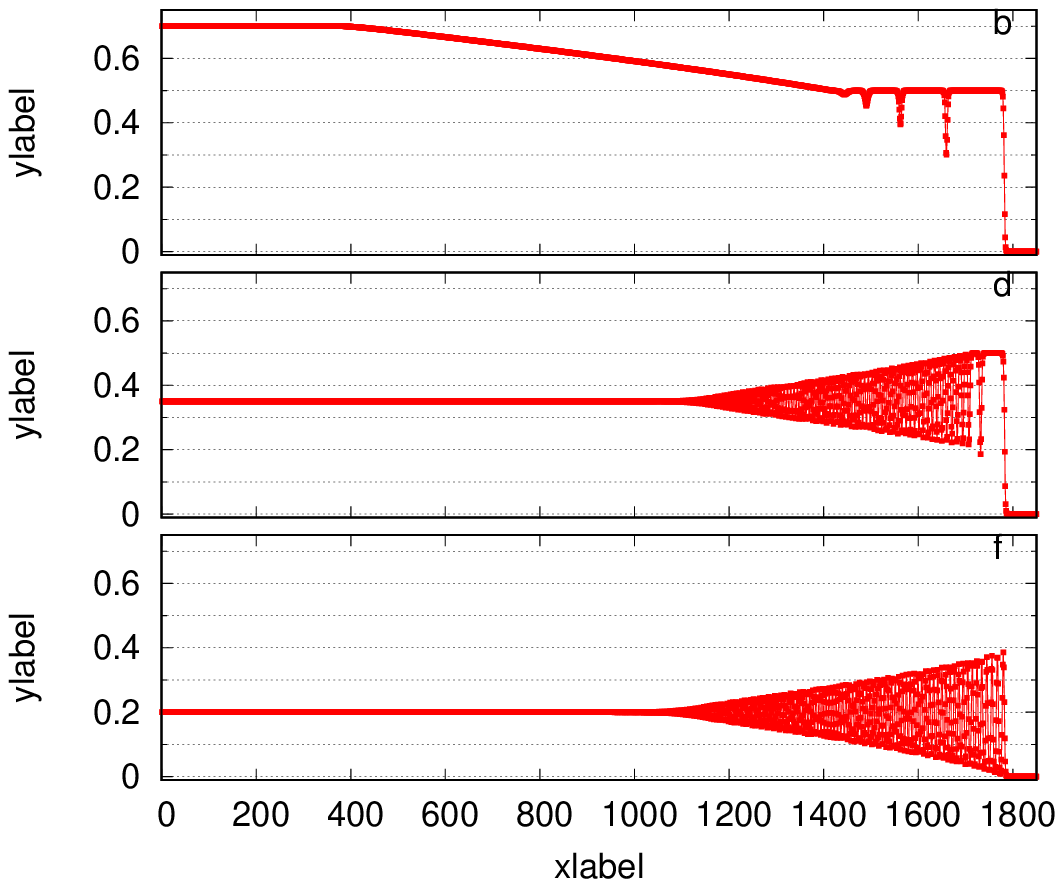}
\caption{ $\alpha=0.25\pi$. Evolution on the lattice for different downstream values of $A$: panels (a-b): $A=0.7\pi$,
panels (c-d): $A=0.35\pi$, panels (e-f): $A=0.2\pi$. The left panel enlarges the vicinity of the front.}
\label{fig:stepev}
\end{figure}

Consider an initial step profile $\theta_n(0)=\frac{A}{2}(1-\tanh \frac{4(n-n_0)}{N_0})$, with $N_0=50$, connecting downstream $A$ with the trivial upstream,
and follow the frontal edge of the propagating wave. Fig.~\ref{fig:stepev}
displays three different  evolution scenario according to order relations between the downstream amplitude $A$ and $\a$.
% We recall that the kink  has an amplitude $2\alpha$ and propagates at the highest admissible speed.

1.  $A>2\alpha$;  as seen in Fig.~\ref{fig:stepev} (a,b), a kink forms. It is followed by  a characteristic linear profile connecting downstream
$A$ with kink's amplitude $2\alpha$. Behind kink's leading  profile few solitons emerge which, due to the invariance \eqref{eq:Invariance},  point down from the top plateau
 $2\alpha$. Being much slower than the kink they lag far behind the emerging dispersionless profile.

2. $\alpha \lesssim A\lesssim 2\alpha$. As seen in Fig.~\ref{fig:stepev} (c,d), a kink forms, but now t is followed by an oscillating domain
with a nearly triangular envelope, which intermediates between the downstream $A$ and kink's frontal amplitude $2\alpha$. The waves-train
can be viewed as a sequence of negative solitons, though in Fig.~\ref{fig:stepev} (c,d) only the first few
pulses have became a truly isolated pointing-down solitons.

3. $ A\lesssim \alpha$ (see Fig.~\ref{fig:stepev} (e,f)). Kink does not form. Instead, there is a wave packet embedded
 within a triangular envelope. In the course of the evolution the leading waves continue to separate from each other to became
true solitons with the leading amplitude $\approx 2 A$ propagating with a velocity as given via $P_{qc}(\lambda;\frac{\pi}{2},2A)=0$.\\

We now append the above phenomenological description of the numerical simulations
with an analysis based on the remarkable proximity we have found between the dynamics on the lattice,
its QC rendition, and the Gardner equation. To this end we shall refer to the  recent  study of the Gardener equation
by Kamchatnov \textit{et al}~[\onlinecite{Kamchatnov-etal-12}] and to an earlier, seminal work of Gurevitch and Pitaevski~[\onlinecite{Gurevich-Pitaevsky-74}]. The comparison is based on their analysis of the signalling problem for Eq.~\eqref{eq:G1}

\begin{equation}
u(x,0) = \begin{cases}
A & \mbox{for } x<0, \\
0 & \mbox{for } x > 0.
\end {cases}
\label{fig:initstep}
\end{equation}

We start noting (the parenthesis refer to the corresponding QC case) that the convection
  $C_{G}=6u(1-u)$ ($C_{cq}(\th)= 2\sin(\frac{\th}{2})\sin(\a-\frac{\th}{2})$) has three key points: $u=0$ ($\th=0$) and $u=1$ ($\th=2\a$)
  where it vanishes, and  a turning point where it attains its  maximal value which separates
  the two domains of monotonicity at $u=1/2~$ in Gardner's case and $~\th=\a$ in the QC.

With the upstream being trivial,  according to Ref.~[\onlinecite{Kamchatnov-etal-12}] there are three regimes according to the position of the downstream amplitude A vs. the three key points:\\
1) { $A<1/2~$ (QC: $A<\a$)},\\
% The downstream and the upstream are on the same side of the monotonicity\\
2) { $1/2<A<1~$ (QC: $\a<A<2\a$)},\\
%The wnstream and the upstream are on the opposite side of the monotonicity\\
3) { $1<A~$ (QC: $2\a<A$)}.\\
We now  detail the dynamics of Gardner's equation vis a vis the numerical results, in parenthesis, in the various regimes.

1){ $A<1/2~$ (QC: $A<\a$).} The downstream and the upstream are on the same side of the monotonicity. In this regime the Gardner equation
 is de facto governed by the KdV equation to which it reduces when the cubic term becomes
 secondary. Consequently, as follows from the analysis in Refs.~[\onlinecite{Gurevich-Pitaevsky-74}] and ~[\onlinecite{Kamchatnov-etal-12}], the  solution takes the
 form of a modulated periodic wave, the so called undular bore, with a lead amplitude having twice its downstream
 value,i,e., $2A$, see Fig.(6) in \cite{Kamchatnov-etal-12}. So much for Gardner; returning to our case,
 panels (e-f) in Fig.~\ref{fig:stepev} clearly show that the analytical features displayed by Gardner's equation parallel {\it the simulation results of the chain}!

2) {$1/2<A<1~$ (QC: $\a<A<2\a$)}. The downstream and the upstream are now on the opposite
sides of convection's monotonicity and Gardner's  solution consists of two parts; let $u(*)=1$ ($\th(*)=2\a$)
be the point where convection vanishes then, provided that $A<u(*)$, instead of a single modular kink
spanning, as in the previous case, the whole upstream-downstream range, we now have a reverse modular kink connecting the
downstream state $A$ with an intermediate state $u(*)$ which then connects to the trivial upstream via a
kink,  see Fig.(8) in \cite{Kamchatnov-etal-12}. Exactly the same phenomenon is seen on panels (c-d)
of Fig.~\ref{fig:stepev} where $\th(*)=2\a$ (that $u(*)=1$ is a consequence of the upstream being trivial.
If $0<u(+\infty)<1/2$,  $u(*)=1-u(+\infty)$ and kink's amplitude depends on $u(+\infty)$ as
well \cite{Kamchatnov-etal-12}).\\

3){  $1<A~$ (QC: $2\a<A$)}. This is a dispersionless regime and rather than a  modulated periodic wave we
have a rarefaction wave which intermediates between the downstream A and a kink at the front, propagating
with the highest admissible velocity,  see Fig.(10) in \cite{Kamchatnov-etal-12}.
Exactly the same effect is observed on panels (a-b) of the chain, with $2\a$ being kink's amplitude.

\subsection{Evolution of an initial pulse}
\begin{figure}[!hbt]
\centering
\psfrag{xlabel}[cc][cc]{space}
\psfrag{ylabel}[cc][cc]{filed $\theta_n/\pi$}
\psfrag{a}[cc][cc]{(a)}
\psfrag{b}[cc][cc]{(b)}
\includegraphics[width=0.45\textwidth]{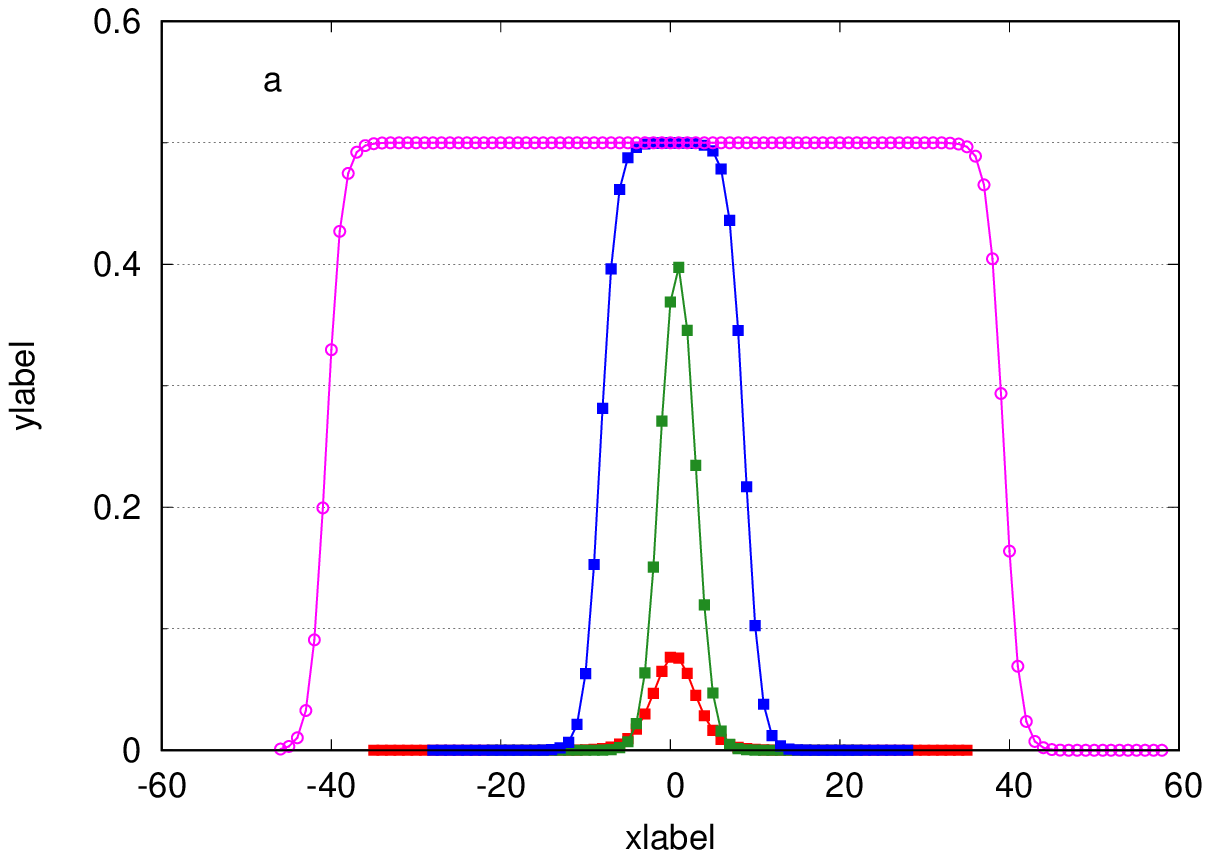}\hfill
\includegraphics[width=0.45\textwidth]{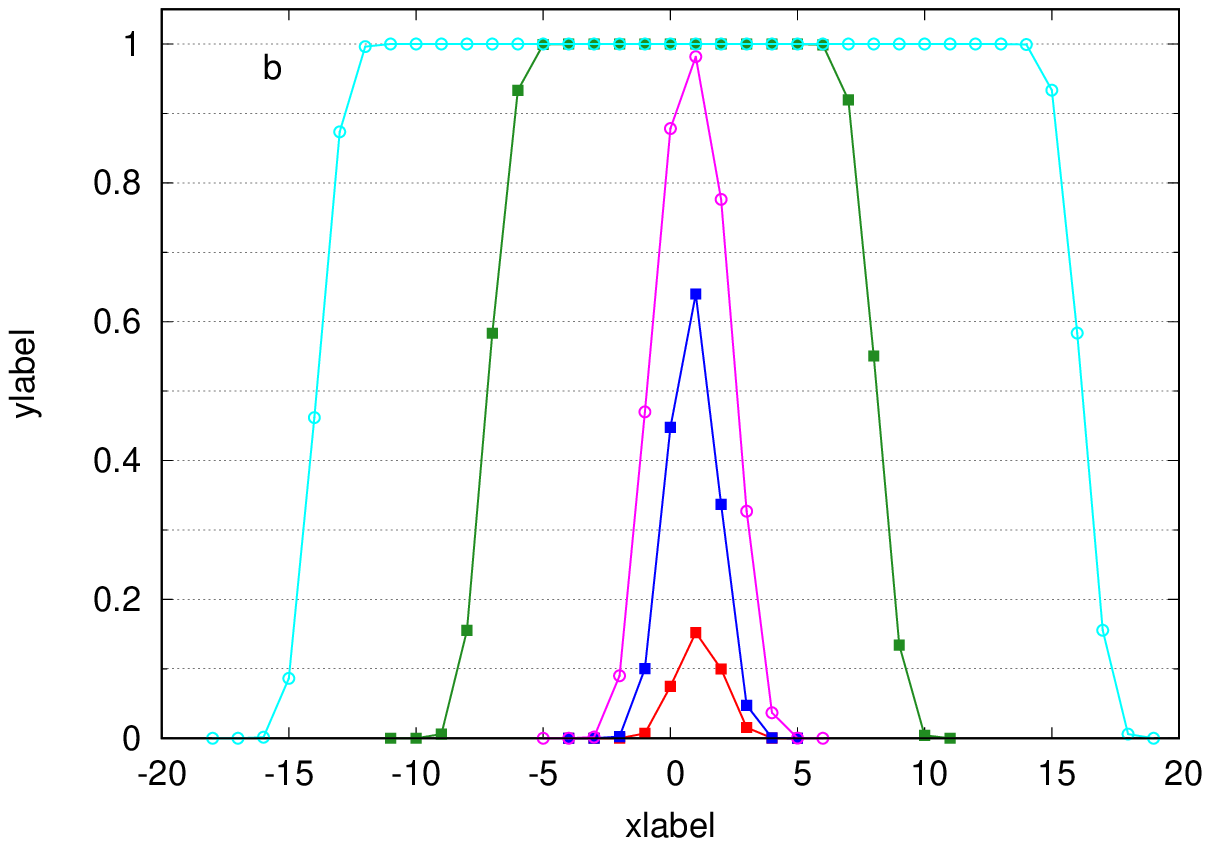}
\caption{Emergence of flatons and kovatons.  Plate (a): $\alpha=0.25\pi$, and from bottom to top:
$A=0.05,\; 0.25,\; 0.5~$ and $1.5$. Plate (b): $\alpha=0.5\pi$, and from bottom to top:
$A=0.05,\; 0.4,\; 0.6,\; 1.0~$ and $1.5$. The difference between the two though 
unobservable to the eye is meaningful; whereas flaton's tails decay exponentially, kovaton's tails decay 
at a doubly-exponential rate which reflects the fact that in the QC limit it has a strictly compact support.}
\label{fig:pev}
\end{figure}

This is arguably the most basic numerical experiment. We follow the evolution of an initially single pulse excitation
 $\theta_n(0)=\frac{A}{\cosh^2 \frac{4(n-n_0)}{N_0}}$, with $N_0=50$. It begets a sequence of solitary waves with
the leading waves being, for small $A$, solitons  which turn into flatons for large $A$'s, see left plate on Fig.~\ref{fig:pev} and kovatons on the right plate.
Though in these simulations flatons and kovatons emerge very naturally,  emergence  of several flatons or kovatons was never observed,
with the solitons forming behind the leading flaton/kovaton having amplitudes smaller than $2\alpha$ and thus slower as well. Notably, the Gardner equation which served us so well in the signalling problem, does not beget flatons easily
 in a corresponding numerical experiments~\cite{ro-or}. They seem to have a very narrow domain of attraction and for a flaton to emerge a special 'tailoring' of initial data was necessary, though once present they have all the features of an integrable entity.

%\begin{figure}[!hbt]
%\centering
%\psfrag{xlabel}[cc][cc]{space}
%\psfrag{ylabel}[cc][cc]{filed $\theta_n/\pi$}
%\includegraphics[width=0.5\textwidth]{fig6.eps}
%\caption{Case $\alpha=0.25\pi$, and different $A$. From bottom to top:
%$A=0.05,\; 0.25,\; 0.5,\; 1.5$.}
%\label{fig:pev}
%\end{figure}

\subsection{Interaction of solitary waves}
\begin{figure}[!hbt]
\centering
\psfrag{xlabel}[cc][cc]{space}
\psfrag{ylabel}[cc][cc]{filed $\theta_n/\pi$}
\psfrag{a}[cc][cc]{(a)}
\psfrag{b}[cc][cc]{(b)}
\includegraphics[width=0.48\textwidth]{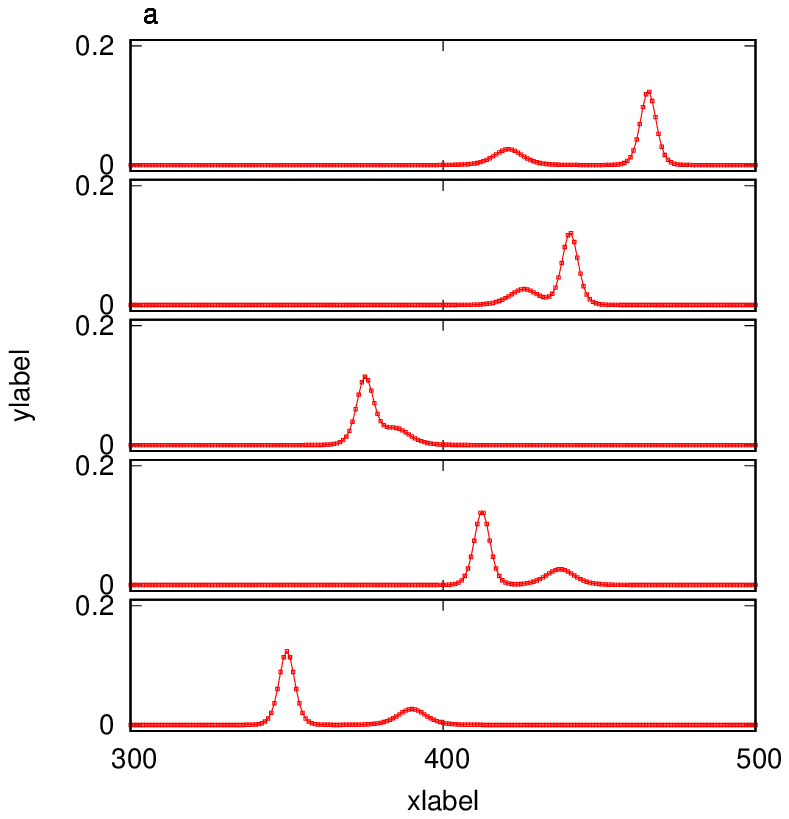}\hfill
\includegraphics[width=0.48\textwidth]{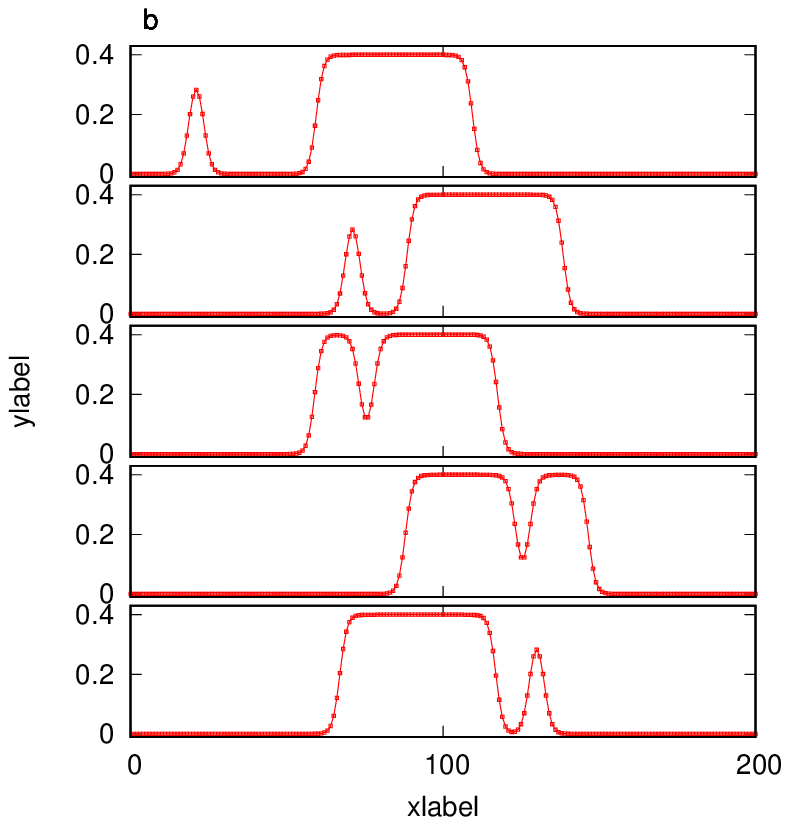}
\caption{Interaction of waves for $\alpha=0.2\pi~$. Panel (a): Interaction of two solitons.
Panel (b): interaction of a flaton with a soliton. The interval between
the frames (starting from bottom) is $\Delta t=150$. The profiles are arbitrarily shifted in $n$-direction.}
\label{fig:coll}
\end{figure}

The essence of our findings is summarized in Fig.~\ref{fig:coll} which
displays collision of two solitons and a collision of a flaton with a soliton: both in Fig.~\ref{fig:coll} 
and other numerical experiments we have carried, the interaction on the lattice of solitary waves, 
whether solitons or flatons is remarkably clean and the interacting entities reemerge 
without  visible distortion or radiation.

\section{Summary}
In this paper we have explored emergence and interaction of nonlinear traveling waves in a phase oscillators chain. We have
found, both for the quasi-continuous rendition and its discrete antecedent, variety of soliton and kink solutions. Furthermore, we
have found that solitons in a velocity range close to kinks flatten and became very wide.  We thus refer to such solitons as flatons.
In 1-D flatons can be looked upon as a joined pair of two kinks. \emph{ In a direct numerical simulations of the chain we have seen the flatons emerging out of
 variety of large initial excitations}. Notably, interaction of solitons and solitons with a flaton are
very clean and without a noticeable distortion or radiation. \\

Finally, we reiterate the  both remarkable role of the Gardner equation~\eqref{eq:G1}, which was deduced at a second stage of approximation of the
 chain, in unfolding the various facets of the dynamics and the actual affinity between  its patterns and the patterns both observed on the chain and predicted by its QC rendition.

\acknowledgments
The authors thank M. Matheny and L. Smirnov for useful discussions.
AP is supported in part by the Laboratory of Dynamical Systems and Applications NRU HSE,
of the Ministry of Science and Higher Education of Russian Federation, grant ag. N 075-15-2019-1931.

%the tree of wisdom for blessing  us with fruitful discussions.
%Merlin.mbs v4.21 2009-07-09.
%

% \bibliography{nld-old,nld-current,%
% pap-ab,pap-ce,pap-fg,pap-hj,pap-kl,pap-mn,pap-oq,pap-rs,pap-tz,%
% pik,books,n-stand}

\end{document}